\documentclass[11pt]{article}
\usepackage{setspace}
\usepackage{amssymb}
\usepackage{amsmath}
\usepackage{amsfonts}
\usepackage{slashed}

\def\be{\begin{equation}}

\def\ee{\end{equation}}

\def\bea{\begin{eqnarray}}

\def\eea{\end{eqnarray}}

\makeatletter
\def\blfootnote{\xdef\@thefnmark{}\@footnotetext}
\makeatother

\begin{document}

\singlespace

\begin{flushright} BRX TH-6693 \\
CALT-TH 2021-031
\end{flushright}

\vspace*{.3in}

\begin{center}

{\Large\bf General Relativity's energy and positivity --- a brief history }

{\large S.\ Deser, FMRS \\}{\it 
Walter Burke Institute for Theoretical Physics, \\
California Institute of Technology, Pasadena, CA 91125; \\
Physics Department,  Brandeis University, Waltham, MA 02454 \\
{\tt deser@brandeis.edu}
}

{\large For Tom Kibble, great physicist and staunch friend, in Memoriam.}

\end{center}

\begin{abstract}
I give a brief review of the search for a proper definition of energy in General Relativity (GR), a far from trivial quest, which was only completed after four and a half decades. The equally (or perhaps more) difficult task of establishing its positivity --- it was to take another fifteen plus years --- will then be  summarized. Extension to cosmological GR is included. Mention is made of some recent offshoots.

An invitation to submit a review to the Proceedings of the Royal Society prompts revisiting a subject of central importance both to GR and to my own past research --- its energy definition and positivity. While there are no loose ends left, a summary may be of some use to students and non-experts.  Exposure to introductory GR is useful.  We will divide this survey into two unequal parts: first, the original GR without a cosmological constant, then extend it to the rather different two cases of $\Lambda \ne 0$.

\end{abstract}

\section{$\Lambda = 0$}

\subsection{Energy definition}

Let's begin with Newtonian gravity whose field is not dynamical, but just a stand-in for action-at-a -distance matter forces: a system's energy is just $E=T+V$ where $V=- G \int\int \rho(r) |r-r'|^{-1} \rho (r')$. The salient fact here is the system's instability: $V$ has no lower limit as its density increases or size decreases, in stark contrast to the GR situation, as we shall see. One other facet of flat space physics
in deep contrast with GR is the following. While Lorentz invariant matter has perfectly well-defined energy, most correctly obtained by first covariantizing it, replacing the flat metric by a here fictitious
$g_{\mu\nu}(x)$, take its action's derivative with respect it, that is as the current responding to a metric variation, then setting it back to flat in the resulting expression for $T^{\mu\nu}(x) = \delta \int dx L(\eta \rightarrow g)/\delta g_{\mu\nu}$, that is defining $T_{\mu\nu}$ as the ``current" response to a metric variation. This is precisely how conserved vector currents are defined for vector coupling, abelian or not: $j^\mu(x) =\delta \int dx L(\partial - i e A) /\delta A_\mu$, whether or not there is an actual gauge field. In both cases, if there is no field, conservation is normal rather than covariant and so conserved ``charges" --- here energy-momentum and angular momentum or electric, but paradoxically, fictitious metric fields suffice to forbid them. There are two ways out, each very deep and relevant to GR: the first is to realize that flat space is defined by the presence of Killing vectors, quantities obeying $D_{(\mu} X_{\nu)}=0$ that turn the $D_\mu T^{\mu\nu} =0$ into an ordinary conserved quantity, $\partial_\mu (T^{\mu\nu} X_\nu)=0$. Let me amplify: In flat space and cartesian coordinates, a matter system's conserved quantities are manifest; however, once generalized coordinates enter, one must use Killing vectors which are of course present irrespective of coordinate choice. In Cartesians, the ten $X_\mu$ are $c_\mu + c_{[\mu\nu]} x^\nu$. The $X^\mu$ are any set of linearly independent vectors that span the space, and the resulting conserved quantities $ \int d^3 x \partial_0(T^{0\mu} X_\mu)$ are the four energy-momenta from $c_\mu$ and the six rotations from the antisymmetric $c_{[\mu \nu]}$. In fancier coordinates, this still holds: for example, each $c_\mu$ ensures that the metric is independent of the corresponding coordinate (from the details of the $\Gamma$ in the now covariant derivative): if there are four independent ones, $g_{\mu\nu}$ is constant, so space is flat as the curvature manifestly vanishes. Each then denotes a Noether symmetry of the first kind, each with its associated constant (of translation) and the $c_{[\mu\nu]}$ the six Lorentz rotations, the total Poincar\'e group. One last point: the matter action in generalized coordinates would seem to have vanishing Hamiltonian, this paradox will best be understood when we turn to GR, although we can note that it represents what is known as the Jacobi form of the action principle --- the whole basis of the ADM analysis!  Briefly, the ADM formulation of General Relativity replaced its original geometric one by a field theoretic basis that exploits its many attributes --- in particular the notion of energy, central to all physics --- as well as existence and properties of gravitational radiation etc.

These essential preliminaries assimilated, we turn to GR itself. [I strongly suggest having a copy of ADM's summary [1] to hand for conventions and details I cannot provide here.] Einstein's greatest insight was not so much that space is curved (that would not be very much different conceptually from its being flat), but that it is dynamical --- a geometry determined both by its sources and as an autonomous system. Built-in to the Riemannian scheme is local coordinate invariance, where the metric is no longer an artifact but the field variable. This gauge invariance of the second kind meant that it was subject to a covariant conservation (Bianchi) identity, rather than on-shell conservation non-identity --- invariance of the first kind with which conserved quantities are associated, as we saw. This means that it is futile to seek an energy definition at this level, as I have also explained elsewhere [2]: one must break the second kind to the first before on can (or should) define Killing vectors, for example. Indeed the august team of Noether and Hilbert foundered on this point when they tried to define energy. Of course, one could not invoke global Killing vectors, for as we saw, they imply flat space without dynamics. One hint --- that those mathematicians were too abstract to note --- is that not all solutions should have a finite, definable energy --- only those with finite excitations, in other words, asymptotically flat at spatial infinity, at all times. This would at least permit Killing vectors out there, but would that be enough to break local in favor of Poincar\'e invariance (we will deal with cosmological GR later)? There are immediately technical questions as to how rapid the falloff must be, a difficult problem in its own right that had to be dealt with. Indeed, an amusing sidelight is that no lesser giants that Einstein and Pauli published a laborious paper showing that a falloff of the metric to Minkowski as fast as $1/r^2$ implied flat space, whereas it seemed clear that anything slower than $1/r$ was too slow. Correct but far from sufficient: for example gravitational radiation (whose existence was also clearly established by ADM; see [1]) with its wavy behavior had to fall more like 
$\hbox{exp}( i(k \cdot x))/r^{3/2+}$, unlike the ``Newtonian" component of the metric with $1/r$ behavior to reduce correctly in the non-relativistic limit. This was indeed a big, but necessary, gap to fill in the search for energy. The technical question was then how to check whether and how the Killings at infinity, where Poincar\'e invariance held for lack of metric excitations, would allow the construction of an energy, in presence of (non-wavy) $1/r$ metric components. What brought it all together was the utterly novel form of the Einstein action as we found it, that we realized had to be so by virtue of general covariance. The paradox here was that its ``Jacobi form" seemed to imply that the Hamiltonian --- let alone the energy --- vanishes! This is worth explaining:  He noticed that the usual system's action
\begin{equation}
I=\int dt L(p,q)= \int [p dq - H(p,q)dt]    \tag{1a}
\end{equation}
could be made covariant by adding an extra degree of freedom, $(P,Q)$ 
\begin{equation}
I= \int [p dq +P dQ- N (P-H) dQ] .  \tag{1b}
\end{equation}
Upon using the $N$-constraint $P=- H$ and the true time, $Q=t$, (1a) is recovered. But due to GR's covariant nature, there is no ``true time" (nor space) coordinate and what therefore emerges is its ``already parametrized" form,
\begin{equation}
I(GR)= \int (dx) [\pi^{ij} \partial g_{ij}/dt - N_\mu R^\mu(\pi, g)]  \tag{2}
\end{equation}
where the four constraint parameters $N_\mu = (g_{0i}, \sqrt{-g}/\sqrt{\, ^3 g})$ are combinations of the $g_{0\mu}$ and the constraints $R^\mu$ depend on the six pair of  variables $(\pi^{ij}, g_{ij})$ where the densities $\pi^{ij}$ are essentially the second fundamental forms, the spatial surface's velocities conjugate to the spatial metric.This leaves as it should, only two degrees of freedom. So no Hamiltonian, nor true time: no initial Eden from which Jacobi expelled us. How do we make a choice then, of this pair? Physically, we do so by looking at the constraints' form at spatial infinity, where we can ``polarize" them into leading, linear, terms plus the rest, as we now show. The $R^\mu$ constraints read
\begin{equation}
\, ^3 R(g) +(\pi^{ij})^2 -(\hbox{tr} \pi)^2 =0,   D_j \pi^{ij}=0  \tag{3}
\end{equation}
where all indices are moved by the --- intrinsically positive --- 3-metric and $R$ is the curvature scalar of   
3-space. These are respectively the energy and momentum constraints already present in the weak field limit. The $\, ^{3}R$ can be polarized into a linear (about flat space) part plus the rest and the former 
provides our energy choice and so too the time. Near infinity, we may decompose the fields according to their, orthogonal under integration, helicity contents --- respectively transverse traceless helicity $2$, vector helicity $1$ and helicity $0$. The latter is the linear part of $\, ^3R$, more precisely $\nabla^2 g^t$, with conjugate time defined by $\pi^t=0$, where conjugate means per the kinetic term in (2). This a (necessary) double miracle: the $\nabla^2 g^t$ means that one can integrate it to be a surface term at infinity AND allows use of the asymptotic Killings as well, in terms of the physical asymptotic falloffs we mentioned. So our first goal --- defining energy --- is reached:
\begin{equation}
\begin{aligned}
E= \int dS_i  \partial_i g^t &= \int d^3x [\sqrt{g} \, ^3R(\hbox{nonlinear}) + (\pi^{ij})^2-\pi^2)/\sqrt{g} ] , \\
 \int d^3 \sqrt{g} \, ^3R(\hbox{nonlinear}) &= \int d^3x \sqrt{g} [\Gamma^k_{li} \Gamma^l_{kj}-\Gamma^l_{lm} \Gamma^m_{ij}] g^{ij}.  
\end{aligned}  \tag{4}
 \end{equation}
The $\Gamma\Gamma$ form can simply be copied from the corresponding one of the 4D Einstein- Hilbert action,
\begin{equation}
I=\int d^4x\sqrt{-g} \, ^4R= \int d^4x\sqrt{-g}[\Gamma^\beta_{\rho\mu} \Gamma^\rho_{\beta\nu} -\Gamma^\beta_{\beta \rho} \Gamma^\rho_{\mu\nu}] g^{\mu\nu} \tag{5}
 \end{equation}
by making all indices there spatial. At the same time, its value as a surface integral at infinity,
per (4) can be read off the coefficient of $1/r$ in $g^t$ as the monopole term in its solution to the (flat space) Poisson equation, $\nabla^2 g^t =$ right hand side of the integrand on the first line of (4). One bit of fine print: strictly speaking, we should have kept track of the asymptotic Killings that are a part of the definition, but that would be too pedantic, since we may simply impose Cartesian coordinates and set each of the $c_\mu= 1$ (at a time) instead. When we come to the cosmological case, the Killings are more complicated, although the formalism is not.   [Note: It is common to use energy and mass interchangeably in the literature, although they are obviously different already in special relativity. The context makes the choice clear.]

\subsection{Positivity of Energy}

The other part of our review deals with the ancient search for a proof (or disproof!) of GR's energy
positivity, an essential attribute for the theory to be stable and thus realistic. I don't think there was
much accomplished here prior to about 1960, when several partial successes were reported [3]. The Einstein-Pauli paper referred to essentially showed (although they did not know this, lacking an understanding of energy) that $E=0$ implies  flat spacetime, i.e., vacuum. This is of course quite important as well, since one would want a physical theory to have just one such minimum and for it to be at vacuum, here flat space. As noted, Noether and Hilbert were off the track because they sought energy in the coordinate invariance, her theorem of the second kind, whereas it is clear that energy is only to be found in her first theorem --- constant transformation invariance, here the global, NOT local, Poincar\'e algebra. This meant in particular breaking the former to the latter by including only asymptotically flat solutions, namely the subset obeying global Poincar\'e, implying the presence of asymptotic Killing vectors there. This also meant that the energy had to be a surface integral at spatial infinity, just like the (also conserved) charges in (abelian or non-)gauge theories, uniquely embodied by the ADM form above.  While its definition was indeed at spatial infinity (GR has no local stress tensor), its evaluation required a volume integration of the very implicit form of its ``right hand side". In particular, unlike any decent matter system, whose Hamiltonian is manifestly positive, e.g., $ \int d^3r [E^2 +B^2]$ for Maxwell or YM, the presence of the GR constraints does not allow this, a rather strange situation!

Some early attempts [3,4] that were a bit too formal, although physically indicative and valid as special cases. One set [3] showed the energy functional's first two variations to be positive, while one fairly general set of positive solutions were physically to be expected: those that enjoyed a moment of pure kinetic excitation (being constant, $E$ stays positive throughout!), in which the ``q" --- spatial metric --- were flat, but only the ``p" --- the $\pi^{ij}$ --- were excited [4]. While the proof is rather simple, it does involve full use of all four constraints as well as of the choice of time conjugate to that of energy; this shows the non-triviality to be expected in the full case. It is instructive to follow: because 3-space is flat, we may decompose any symmetric tensor --- not necessarily small --- by helicities, $2$, $1$ \& $0$ ($2$, $3$ ,$1$ in number respectively). Their amplitudes are orthogonal under integration, that is there are no cross terms in $\int d^3x(\pi^{ij})^2$. Actually, this is not even necessary to use: The helicity $0$, $\pi^t$ part, vanishes as it defines the time choice corresponding to that of energy since they are conjugate. Helicity $1$ vanishes by the constraints $\partial_j \pi^{ij} =0$ (in Cartesian coordinates of flat space), leaving only the transverse-traceless helicity $2$ excitation, with manifestly non-negative energy
\begin{equation}
E= \int d^3x (\pi^{tt})^2,     \tag{6}
\end{equation}
vanishing only at unique flat vacuum, $g_{ij}=0$. On the other hand, the other extreme, vanishing $\pi^{ij}$ 
but non-weak $g_{ij}$, is not easily treated. The general energy expression (4) makes this obvious: the $\int \Gamma\Gamma$ terms are NON-definite, even when expressed in terms of the metrics and use of the single available metric constraint, namely that $g^t$ is the energy ``density".  Another special case of note also represents a very symmetric case [4].

The real proofs differed considerably in complexity. The first was given via Supergravity (SUGRA) [5] whose ``Dirac square root" nature did exhibit a manifestly positive, if rather formal, expression, then by rigorous mathematics [6] and finally spinorially, as inspired by SUGRA --- directly in classical GR [7]. The SUGRA way is amazingly direct: any supersymmetric theory is characterized by conserved spinorial supercharges $Q$ associated with invariance under constant spinorial variations. The SUGRA invariance is of course not under constant variations; instead, they reduce to asymptotic constant spinor transformations at spatial infinity, quite like the constant Poincar\'e invariance and Killing vectors there. That is, GR energy and SUGRA charge are defined for asymptotically flat worlds and are bound by the rather intuitive relation, $E=Q^\dag Q > 0$, for Majorana spinors $Q$, since the spin $3/2$ fields $\Psi_\mu$ are Majorana. Now we descend from necessarily quantized SUGRA to classical pure GR as follows: First, set $h=0$, thereby eliminating closed loops, then remove all graphs with external spin $3/2$ lines, in which case there are none anywhere, as there are no loops. Since $E>0$ is valid fo the full SUGRA, it also holds for this limit of classical GR, QED. The fancy differential geometrical proof involved arcane properties of the curvature not easily summarized in a physicist's review. Finally I add the obvious --- that any normal, i.e., positive energy, matter contributes positively as well, since in covariant form we saw that its energy was simply part of the N-constraint and its energy positivity persists under covariantization!

Since all the above results are quite solid, I permit myself one additional, handwaving, argument for
$E>0$. As mentioned in the introduction, Newtonian gravity is UNstable because the attractive potential 
can grow arbitrarily large and negative as the masses increase or the system's size goes to zero. The difference in GR is, as can be found in [1], that instead of the (unbounded from below) Newtonian $m=m_0 - G m_0^2/r$ where $m_0$ is the mechanical mass and $r$ the size, of a particle, GR says that ALL energy gravitates, that is 
 \begin{equation}
m= m_0 - Gm^2/r,  \tag{7}
\end{equation}
where $m$ is the total mass including the self-interaction energy. Solving the quadratic for $m$ shows that zero is its lower bound; physically nothing is left to gravitate when $m=0$ is reached! The above is indeed quite correct for a massive particle source form of the field equations. Whether the GR field itself obeys a similar constraint seems most likely, but is speculation. Let me elaborate a bit: In (7), take $r$ to be some characteristic size of a (positive energy) weak field distribution $m_0$; that choice is defensible since as the total mass drops to zero and perhaps beyond, it must go through this stage. The solution of (7) is of course 
\begin{equation}
2Gm/r = -1+ \sqrt{1+4Gm_0/r},  \tag{8}
\end{equation}
the upper branch being forced by continuity with the $G \rightarrow 0$  limit, $m\rightarrow m_0$. But this is manifestly positive, as desired.  Here $r$ is the radial coordinate in conformal gauge, but since $E$ is an invariant, the conclusion is unaffected.

\section{$\Lambda \ne 0$}

\subsection{Energy definition}
Because the asymptotics here are the closed, spherical (deSitter$=$dS), or open hyperbolic (AdS) according to the $+/-$ sign of $\Lambda$ (with our conventions), the associated group is no longer Poincar\'e;
this deterred any attempts at defining energy, though clearly the dS/AdS-Schwartzchild solutions
contains a term very like the $m/r$ of the $\Lambda=0$ theory. Indeed, it was not until 1982 that the solution was found [8], in terms of the correct Killing vectors at the ``edges"; it is the only relevant reference to our needs. The dS/AdS situations are very different, as might be expected from their behaviors there. In the dS case, space is closed and our universe ends at an event horizon, while for AdS there is no bound. [Amusingly, the actual universe seems to be dS, while theoretical models, especially supergravity ones, can only be consistent in AdS.]  AdS is, instead, infinite. Specifically, these spaces are invariant under rotations of the corresponding hyperspheres/hyperboloids, $\pm x_0^2 + \ldots +x^4 =c$.  
The corresponding $10$ angular momenta $J_{[\mu\nu]}$ are the equivalents of the $10$ Poincar\'e translations \& rotations; in particular $J_{04}$ is the energy equivalent, to which it reduces at $\Lambda=0$. Again, there are asymptotic Killings vectors. The added cosmological term, $\Lambda \sqrt{-g} = N\sqrt{g}$, only adds a constant to the $N$-constraint (quite an only, though, since it alters the leading term!). There is again a dominant linear Coulomb-like term in the energy constraint, but now the spaces are very different: dS has an intrinsic horizon within which we live and can assess positivity of excitations.

\subsection{Positivity}

The dS case can be well handled for weak excitations. Stability is easily established for small fluctuations about the de Sitter background, occurring inside the event horizon and semiclassical stability analyzed. Absent supersymmetry, this is perhaps the best possible result, though the mathematical proof of [6] might be pursued here. For $\Lambda < 0$, positivity and stability are demonstrated for all asymptotically anti-de Sitter metrics. The analysis is based on the general construction of conserved flux-integral expressions associated with the symmetries of a chosen background. It is almost trivially proved, using the above supergravity techniques for SUGRA, in terms of the graded anti-de Sitter algebra with spinorial charges also expressed as flux integrals, that of the SUGRA case plus the extensions to hyperbolic asymptotics, yielding once again the energy as the (positive) square of the supercharge.

\section{Recent developments}
Once the basics were established, progress continued along two broad lines: Energy of extended gravity models, and other possible constants. The first involves extending the notion of energy to higher curvature --- such as Weyl's conformally invariant and other quadratic (and higher powers) --- actions (whose energies are not, in general, positive nor vanishing - as erroneously stated by  some authors for Weyl gravity) see, e.g., [9], 
and to other dimensions, in particular to $D=3$'s topologically massive gravity. The second strand concerns null, rather than spatial infinity invariances, such as the currently fashionable Bondi-Metzner-Sachs (BMS) ``supertranslation" group. As it involves quite different concepts, it is better left to separate reviews.  Finally, some attempts at establishing positivity of alternate energy definitions failed to do so, e.g., [10].

\section*{Acknowledgements}
This work was supported by the U.S.Department of Energy, Office of Science, Office of High Energy Physics, under Award Number de-sc0011632. I thank J.\ Franklin for helpful Latexing.

\end{document}